\documentclass[12pt]{article}
\usepackage{mathpple}
\usepackage{palatino,graphicx,cite,epsfig,fancyheadings,setspace}
\usepackage[letterpaper=true,pdftitle={Telomere cancer}]{hyperref}

\begin{document}
\pagestyle{fancy}


\section*{Telomere loss limits the rate of human epithelial tumor formation.}


\begin{centering}
Hermann B. Frieboes and James P. Brody\\ 
\emph{Chao Family Comprehensive Cancer\\
and\\
Department of Biomedical Engineering\\ 
University of California--Irvine\\ 
Irvine, CA 92697-2715\\}
\end{centering}


\begin{quote}
\textbf{
Most human carcinomas exhibit telomere abnormalities early in the
carcinogenesis process \cite{meeker2002,vanheek2002,meeker2004}
suggesting that crisis caused by telomere shortening may be a
necessary event leading to human carcinomas
\cite{artandi2000}. Epidemiological records of the age at which each
patient in a population develops carcinoma are known as age-incidence
data; these provide a quantitative measure of human tumor initiation
and dynamics \cite{knudson2001}. If crisis brought on by telomere
shortening is necessary for most human carcinomas, it may also be the
rate limiting step.
To test this, we compared a mathematical model in which telomere loss
is the rate limiting step during carcinogenesis with age-incidence
data compiled by the Surveillance, Epidemiology and End Results (SEER)
program \cite{seerstat}.
We found that this model adequately explains the age-incidence data.  
The model also implies that two distinct paths exist for carcinoma to
develop in prostate, breast, and ovary tissues.
We conclude that a single step, crisis brought on by telomere
shortening, limits the rate of formation of human carcinomas. }

\end{quote}

\newpage

\subsection*{Introduction}


Crisis brought on by shortening telomeres, or senescence, may be a
necessary step in epithelial carcinogenesis.  A model of cellular
senescence suggests that cells reach their replicative limit, the
Hayflick limit \cite{hayflick1965}, as the length of their telomeres
shorten \cite{shay2000,bodnar1998}.  Upon reaching this limit, cells
face a crisis. A few cells, probably those that have acquired certain
mutations, are able to overcome this crisis and continue replicating
\cite{wright1989}, while most others cannot survive and perish.  The
surviving cells, replicating without functional telomeres, exhibit
massive genomic instability.  Genomic instability is also
characteristic of epithelial cancers \cite{artandi2000}. The
epithelial carcinogenesis process exhibits telomere abnormalities both
early and often \cite{meeker2002,vanheek2002,meeker2004}. Hence,
crisis caused by telomere shortening may be a necessary event in
epithelial carcinogenesis.




If crisis is necessary for epithelial carcinogenesis, it may also be
the rate limiting step. Identifying the rate-limiting step is key to
delaying the onset of carcinogenesis.  If one reduces factors that
are not rate-limiting, no significant change in onset will result.  

Surveillance networks record the age at which carcinoma is diagnosed
in patients.  These cover just over 25\% of the United States
population.  With census data, this can be converted to a rate.  We
use this data to test the hypothesis that crisis brought upon by
reaching cellular senescence is the rate limiting step in carcinoma
formation.


Long ago \cite{nordling1953,armitage1954}, it was pointed out that if
$r$ mutations are required before carcinoma develops, if each mutation
occurs at a constant rate, and if cells with less than $r$ mutations
have no growth advantage, then the age-specific incidence of carcinoma
should be $I(t)=kt^{r-1}$ \cite{knudson2001}.  Although, this model
has evolved \cite{armitage1957,moolgavkar1981} to include effects such
as an intermediate growth advantage, its basic conclusion
remains. When this model is applied to colon carcinoma age-incidence
data collected by cancer registries, it leads to the conclusion that
four to six mutations are required to transform a normal cell into a
malignant one.  This conclusion is now widely accepted
\cite{hanahan2000a}.

However, this model fails to explain several anomalies with the
age-incidence data.  Breast carcinoma incidence data does
not follow this model.  (This has been explained by a model
suggesting that breast tissue ages at a different rate than calendar
time \cite{pike1983}).  Prostate carcinoma incidence increase much
more rapidly with age than most others, implying that 20 to 30
mutations are required for transformation.  Also, the rate of increase
in the incidence for several carcinomas begins to slow or even drop at
advanced ages \cite{pompeii2002}.  None of these phenomena can be
explained with the current model.



Widespread biochemical and medical imaging screening programs
implemented in the United States during the past two decades have led
to earlier and more complete diagnosis of carcinoma.  This, in turn,
has led to dramatic differences in the age-incidence data collected by
cancer registries.  This data presents an increasingly more accurate
picture of when carcinoma actually develops, as opposed to when it
becomes fatal.

\subsection*{Methods}

\textbf{Mathematical Model} We constructed a mathematical model to
describe the expected age-incidence of carcinoma.  The mathematical
model is based upon the biological model that carcinoma only develops
after the onset of crisis.  The assumption is that epithelial
carcinogenesis has a single rate limiting step and that step occurs
when telomere shortening has reached its limit.  The
simplest model would be that carcinoma develops in every member of the
population at exactly the same age.  However, due to the heterogeneous
population living in a heterogeneous environment, we would expect that
this statement generalizes to the mathematical statement that the
age-incidence follows a normal distribution.  

Although it is not possible to rigorously justify its use here, the
normal distribution is ubiquitous in nature \cite{denny2000}.  Its
finite value at negative ages may present theoretical objections to
its application in this instance, but as long as the mean is large
enough and standard deviation small enough (which is true in all
cases here), these theoretical objections have no practical grounding.


The age-incidence, $I(t)$, of carcinoma for the model is given by
\begin{equation}
I(t)=\alpha \left(\frac{N(t,\tau,\sigma)}{1-\int _0 ^t N(s,\tau,\sigma)ds}\right), 
\label{single}
\end{equation}
where $\alpha$ represents the number of people susceptible to the
cancer.  $N(t,\tau,\sigma)$ is the well-known normal distribution,
also called the Gaussian distribution, or bell-shaped curve. It is
given by
\begin{equation}
N(t,\tau,\sigma) =  \frac{1}{\sqrt{2\pi \sigma^2}} {\rm e}^{-\left(\frac{(t-\tau)^2}{2\sigma^2}\right)} .
\label{gaussian}
\end{equation}
The factor in the denominator of Equation\,\ref{single}, $
1-\int_0^tN(s,\tau,\sigma)ds$, adjusts for the number of people who
already have acquired carcinoma.  For most carcinomas, it is very
close to one. It ranges from 1.0 to 0.94 for colon carcinoma.

If carcinoma can develop by two distinct paths,
the observed age-incidence will be a linear combination of two
independent functions,
\begin{equation}
I(t)= 
\alpha_1 \left( \frac{N(t,\tau_1,\sigma_1)}
                     {1-\int _0 ^t N(s,\tau,\sigma)ds} 
	\right) + 
\alpha_2 \left( \frac{N(t,\tau_2,\sigma_2)}
                     {1-\int _0 ^t N(s,\tau,\sigma)ds} 
        \right) .
\label{double}
\end{equation}

\textbf{SEER Data} We tested the models, Equation~\ref{single} and
Equation~\ref{double}, by fitting them to age-incidence data from
different carcinomas.  The age-incidence data were recorded by the
SEER 12 registries~\cite{seerstat} and the rates were age-adjusted to
the 2000 US population as measured by the census. The SEER registries
have compiled cancer incidence information on a large representative
sub-population of US residents since 1975.  From this database, we
selected patients diagnosed in 2000 with carcinoma in the indicated
tissue.  This excludes the small number with other types of cancers,
sarcomas for instance, which probably arise through a different
process.  The calculation of confidence intervals are based upon the
method of Fay and Feuer \cite{fay1997}.

We chose to analyze data for patients who were diagnosed in the same
year (2000), rather than those born in the same year (a birth-cohort).
Different factors could distort either data set.  Birth-cohort
analysis is significantly distorted by changes in medical practice and
diagnostic technology. On the other hand, changes in environmental
carcinogens may distort the data presented here.  Detection technology
for the carcinomas presented here have dramatically improved over the
past 50 years.  Hence, we focus on patients diagnosed in a single
year, while recognizing that their environmental exposure may be
different.


\textbf{Model Fitting} The model was fit to the data by minimizing
chi-squared, $\chi^2$ \cite{taylor1997},
\begin{equation}
\chi^2=\sum_{k=1}^{17}\frac{(O_k-E_k)^2}{E_k},
\end{equation}
where $O_k$ is the observed number of cases of cancer and $E_k$ is the
expected number of cases of cancer in each of the 17 age ranges. The
18th age group includes all those diagnosed with cancer at or above 85
years of age.  This age group was not included in calculating the
minimization, since it is unbounded.  The expected number of cases, $E_k$, was
obtained by multiplying the age-corrected incidence function $I(t)$,
evaluated at the mid-point of the age group range, by the population
under surveillance in that age range.  The parameters $\alpha$, $\tau$,
and $\sigma$ were varied to minimize $\chi^2$. This optimization used
the Generalized Reduced Gradient algorithm \cite{lasdon1978} to
perform the minimization.  Estimates on the errors of the parameters
were obtained by fitting five independent data sets (data from five
different years, 1996 -- 2000) and reporting the mean and standard
error of these measurements.

\subsection*{Results}
\begin{figure*}[htb]
\begin{center}
  \includegraphics[width=5in]{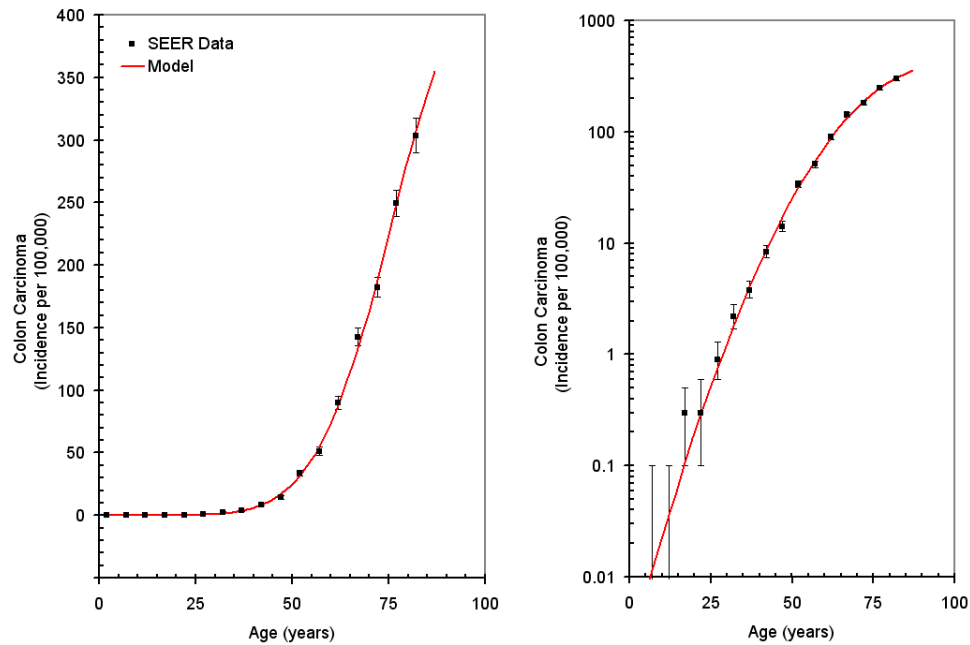}
\end{center}

\caption[]{The observed age-incidence of colon carcinoma (A) and
breast carcinoma (B) compared to the model.  The left and
right panels show the same data.  On the left the incidence is plotted
on a linear scale, while on the right it is plotted on a logarithmic
scale.  The logarithmic scale better represents lower incidence levels,
while the linear scale better represents higher incidence levels.  In
each case, the measured incidence is represented by a point and the
95\% confidence intervals by error bars.  The solid lines represents
predicted incidence levels based upon the model. }

\label{colon2000}
\end{figure*}

Age-incidence data from carcinomas originating in the colon, stomach,
pancreas, and esophagus are consistent with the single path model,
Equation~\ref{single}.  In each case, we found an excellent fit
between the model and the actual age-incidence data, see
Figure~\ref{colon2000} for the fit to colon carcinoma data, and the
Supplemental Figures for the others.

\begin{figure}[htb]
\begin{center}
  \includegraphics[width=6in]{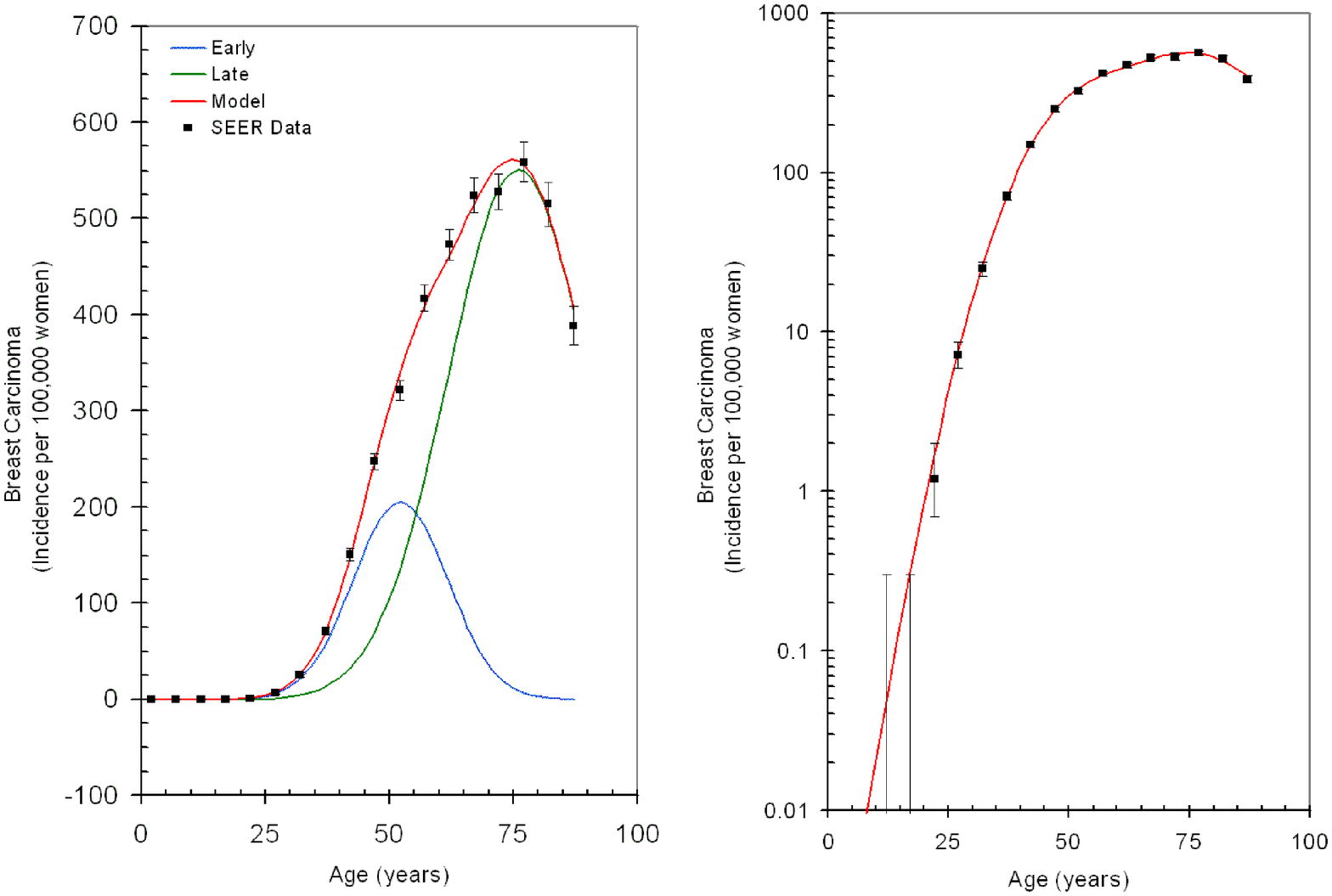}
\end{center}
\caption[]{The age-incidence of breast carcinoma. As in Figure 2, the
left and right panels show the same data.  On the left the incidence
is plotted on a linear scale, while on the right it is plotted on a
logarithmic scale.  In each case, the measured incidence is
represented by a point and the 95\% confidence intervals by error
bars.  The solid lines represents predicted incidence levels based
upon the model.  The age-incidence data can only be explained if two
distinct tumorigenesis paths exist.  One, the early-onset path, is
represented by the blue line, and the other, the later-onset path, by
the green line.  In this case, the predicted incidence is the sum of
the two, represented by the red line. }
\label{breast2000}
\end{figure}

Breast, ovarian, and prostate carcinoma are consistent with the
existence of two paths in the model, Equation~\ref{double}.  In
ovarian carcinoma, see the Supplemental Figures, these two paths have
dramatically different mean times, 22 years and 78 years, and the
early version is over a hundred times rarer.  While in prostate
carcinoma, see the Supplemental Figures, the two mean times are
relatively close, 68 and 80 years.  In breast carcinoma, see
Figure~\ref{breast2000}, the two versions of the disease occur far enough
apart, 52 and 76 years and the population-at-risk is similar enough,
5\% and 19\%, so that the two populations can be discerned. We also
examined other non-epithelial cancers.  These generally are not
consistent with the presented model.  In Table~1, we present the
fitted parameters for each carcinoma along with error estimates.

\begin{table}[h]
\caption{Three parameters describe the age-incidence curve for each
carcinoma.  The first parameter, $\alpha$, represents the fraction of
the population susceptible to the disease. In the case of ovarian and
breast cancer it is the fraction of women; while for prostate
carcinoma it represents the fraction of men.  The second parameter,
$\tau$, is the mean time (in years) to develop the
disease. The final parameter, $\sigma$, is the standard deviation of
the time (in years).}

\label{parametertable}
\begin{center}
\begin{tabular}{llp{1.5in}p{1.5in}p{1.5in}}\hline\hline
Tissue&&Susceptible        population ($\alpha$)&Mean time ($\tau$) (years)&S.D. time ($\sigma$) (years)\\\hline
Colon&&$20.0(\pm1.3)$\%&$90.8(\pm1.0)$&$18.8(\pm0.3)$\\
Stomach&&$~5.2(\pm0.7)$\%&$101.0(\pm2.6)$ &$22.2(\pm0.8)$\\
Pancreas&&$~3.2(\pm0.2)$\%&$87(\pm1.6)$&$17.1(\pm0.7)$\\
Esophagus&&$~1.0(\pm0.1)$\%&$80(\pm2.8)$&$15.3(\pm1.1)$\\
Breast &Early&$~5.1(\pm0.7)$\%&$52.0(\pm0.5)$&$9.4(\pm0.1)$\\
       &Late&$17.2(\pm1.4)$\%&$75.3(\pm0.6)$&$13.4(\pm0.9)$\\
Prostate&Early&$~12(\pm1)$\%&$67(\pm1)$&$8.4(\pm0.2)$\\
        &Late&$~16(\pm1)$\%&$79(\pm1)$ &$9.4(\pm1.3)$\\
Ovary&Early&$~0.02(\pm0.004)$\%&$23(\pm1.3)$&$5.4(\pm0.7)$\\
        &Late&$~3.1(\pm0.3)$\%&$78(\pm2.4)$ &$21.1(\pm1.1)$\\\hline\hline

\end{tabular}

\end{center}
\end{table}

\begin{figure}[htb]
\begin{center}
  \includegraphics[width=6in]{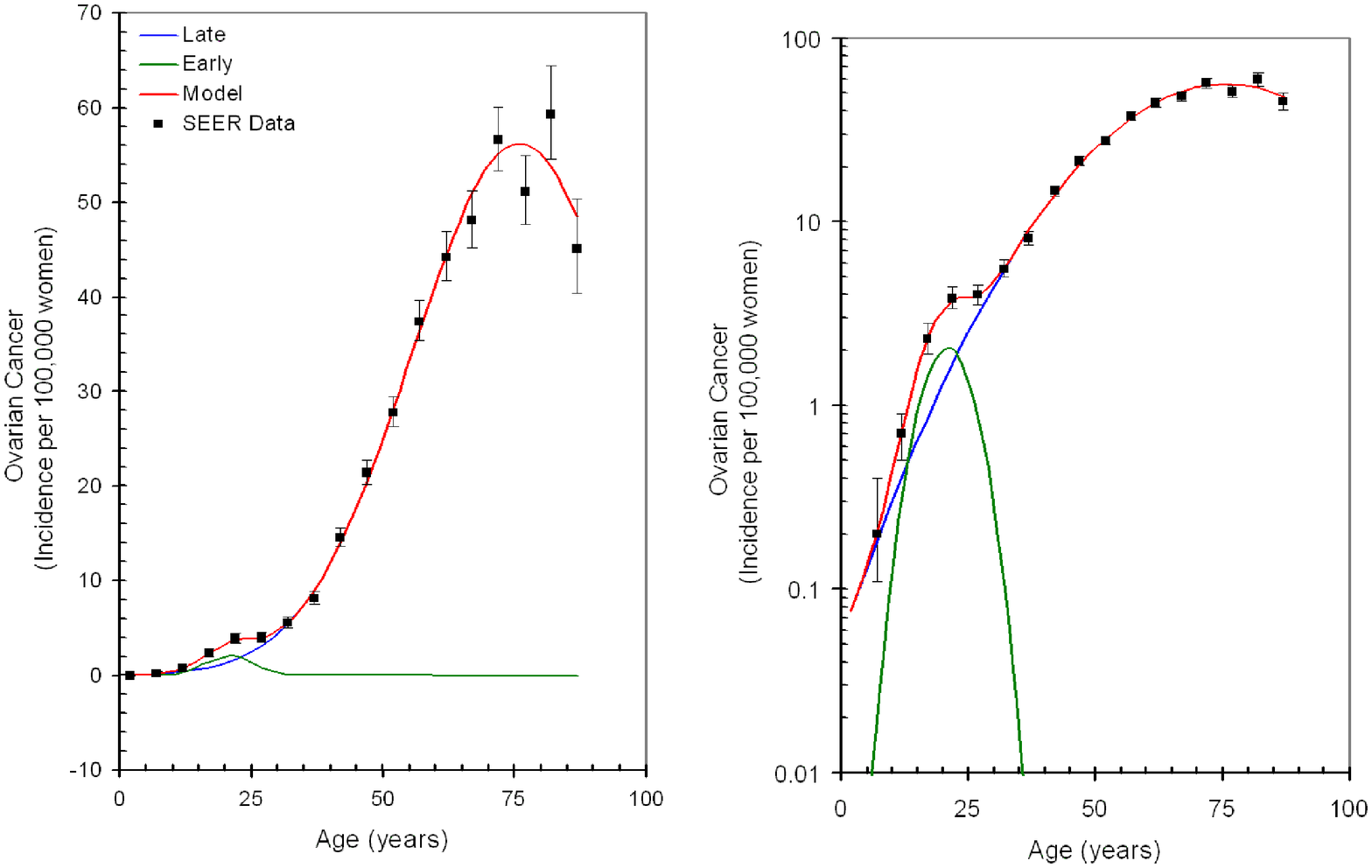}
\end{center}
\caption[]{The age-incidence of ovarian cancer. As in Figure 2,
the left and right panels show the same data.  On the left the
incidence is plotted on a linear scale, while on the right it is
plotted on a logarithmic scale.    In each case, the measured
incidence is represented by a point and the 95\% confidence intervals
by error bars.  The solid lines represents predicted incidence levels
based upon the  model. }
\label{ovary2000}
\end{figure}


\subsection*{Discussion}

The results presented here indicate that the age-incidence data is
consistent with a model in which cellular senescence is the
rate-limiting step in the development of carcinoma.

The mathematical model has three parameters, and each has a well
defined meaning. The first parameter, the mean time, $\tau$, indicates
the average time, measured from birth, for the formation and detection
of the cancer.  One would expect that environmental influences can
affect this parameter, and that may be one explanation for the
variation of times observed in the population. The second parameter,
the standard deviation of the time, $\sigma$, quantifies this
variation in the population.  This variation can be attributed to
either intrinsically random processes, genetic variation, or
environmental differences within the population.  The final parameter,
the susceptible population, $\alpha$, describes the fraction of the
total population susceptible to this carcinoma.  This parameter is
always greater than the actual fraction who develop the disease, since
some may die before developing the disease.

Two surprising conclusions arise from this analysis. First, two
distinct paths exist for the development of breast, ovarian, and
prostate carcinomas. Second, substantially less than 100\% of the
population is susceptible to developing any specific carcinoma.

Other studies also suggest that two distinct breast carcinomas exist.
Early-onset breast carcinoma is already a recognized subclass of the
disease, typically being described as occurring in women before the
age of 35 \cite{walker1996}. Our results indicate significant overlap
between the early and late onset forms, and age itself is insufficient
to determine whether a woman has one form of the disease or another.
However, most cases (about 80\%) diagnosed before the age of 35 will
be early-onset.  Published work concludes that early-onset disease is
biologically distinct \cite{johnson2002,weber-mangal2003} from the
late onset version and also a more potent form of the disease
\cite{chung1996}.  Hence, our conclusion is consistent with the modern
view of breast cancer.

An alternative explanation for the atypical age-incidence curve of
breast carcinoma has been proposed \cite{pike1983}.  This explanation
suggests that breast tissue does not age linearly, but rather as a
function of hormonal levels.  If this ``breast age'' is properly
considered, the age-incidence of breast cancer more closely follows
that typically seen in other cancers.  This is the currently accepted
view \cite{easton2000}.  In contrast, our model explains the breast
cancer age-incidence based on two widely accepted concepts: genomic
instability, and the presence of two distinct forms of the disease.


The age-incidence data, viewed in the context of this mathematical
model, implies that only a subset of the population is susceptible to
developing the disease. Similar conclusions have been drawn by others.
Peto and Mack \cite{peto2000} based their conclusion on a study of the
incidence of breast cancer in monozygotic twins.  A similar study
\cite{hemminki2002} performed on a more extensive dataset suggests
that this result is an artifact. However, the conclusion is supported
by a segregation analysis \cite{pharoah2002} and by this
work. Thilly's group \cite{herrero-jimenez1998,herrero-jimenez2000}
has extended the Nordling, Armitage-Doll model to include the
processes of mutation, cell growth and turnover while also accounting
for heterogeneity of both genetic factors and environmental exposure.
They concluded that the sub-population at risk for colon carcinoma was
about 42\% and has been invariant for over a century.

Different mechanisms could explain the existence of a minority that is
susceptible to carcinoma.  For instance, members of this minority may
have inherited susceptibility.  Although we can rule this out for
high-penetrance alleles, since by definition sporadic carcinoma does
not show familial aggregation, it is thought that low-penetrance
alleles \cite{houlston2004} may be responsible for this.  An alternative
is that membership in this minority may indicate the acquisition of a
somatic mutation early in life as proposed by
Frank and Nowak \cite{frank2003}.  They reason that most carcinomas
could result from somatic mutations occurring within stem cells during
development. Finally, it may be an indication of This mechanism is consistent with our findings.

Although some forms of carcinoma are clearly associated with
environmental exposure, the mechanism causing this has not been
clearly determined.  The most widely held hypothesis, that
environmental carcinogens induce point mutations thus causing cancer,
is not supported by observation \cite{thilly2003}.  The analysis
presented here suggests an alternative. If environmental carcinogens
cause more rapid cellular proliferation, the Hayflick limit would be
reached at a younger age, and carcinoma could develop earlier. Thus
environmental carcinogens would lead to increased rates of carcinoma.

The conclusions here are based on the validity of the SEER data. This
is the most comprehensive dataset available.  Audits are regularly
conducted to ensure the completeness and accuracy of the dataset.
However, others \cite{hertzpicciotto2001} have pointed out that the
SEER population numbers do not exclude those patients who cannot
contract the disease.  For instance, women who have had hysterectomies
cannot contract ovarian cancer.  Approximately 0.5\% of women have
hysterectomies annually \cite{farquhar2002}.  Thus, the ovarian
carcinoma incidence-rate is likely to be systematically underestimated
in the SEER data, particularly in the elderly.  We estimate this
systematic error to be as large as 20-30\%.  Although this would
affect the numbers reported in Table~1, it is unlikely to affect the
overall conclusions.  Hysterectomies are the most common surgical
procedure, hence we expect systematic errors due to this effect to be
much less in other carcinomas.

In conclusion, we found that the formation of human carcinomas is
limited by a single step, consistent with a model in which crisis
brought on by telomere shortening is the key step. Furthermore, if
this is true, then carcinomas predominantly develop in a susceptible
minority of the population and carcinoma of the breast, ovary,
and prostate can occur through two distinct paths.

\subsection*{Acknowledgments}

This work was supported by grant HG-00047 from the National Institutes
of Health. We thank Steven Frank for useful discussions.

\clearpage


\bibliographystyle{unsrt}

\bibliography{../brody}
\markboth{}{}

\newpage

\begin{figure}[htb]
\begin{center}
  \includegraphics[width=6in]{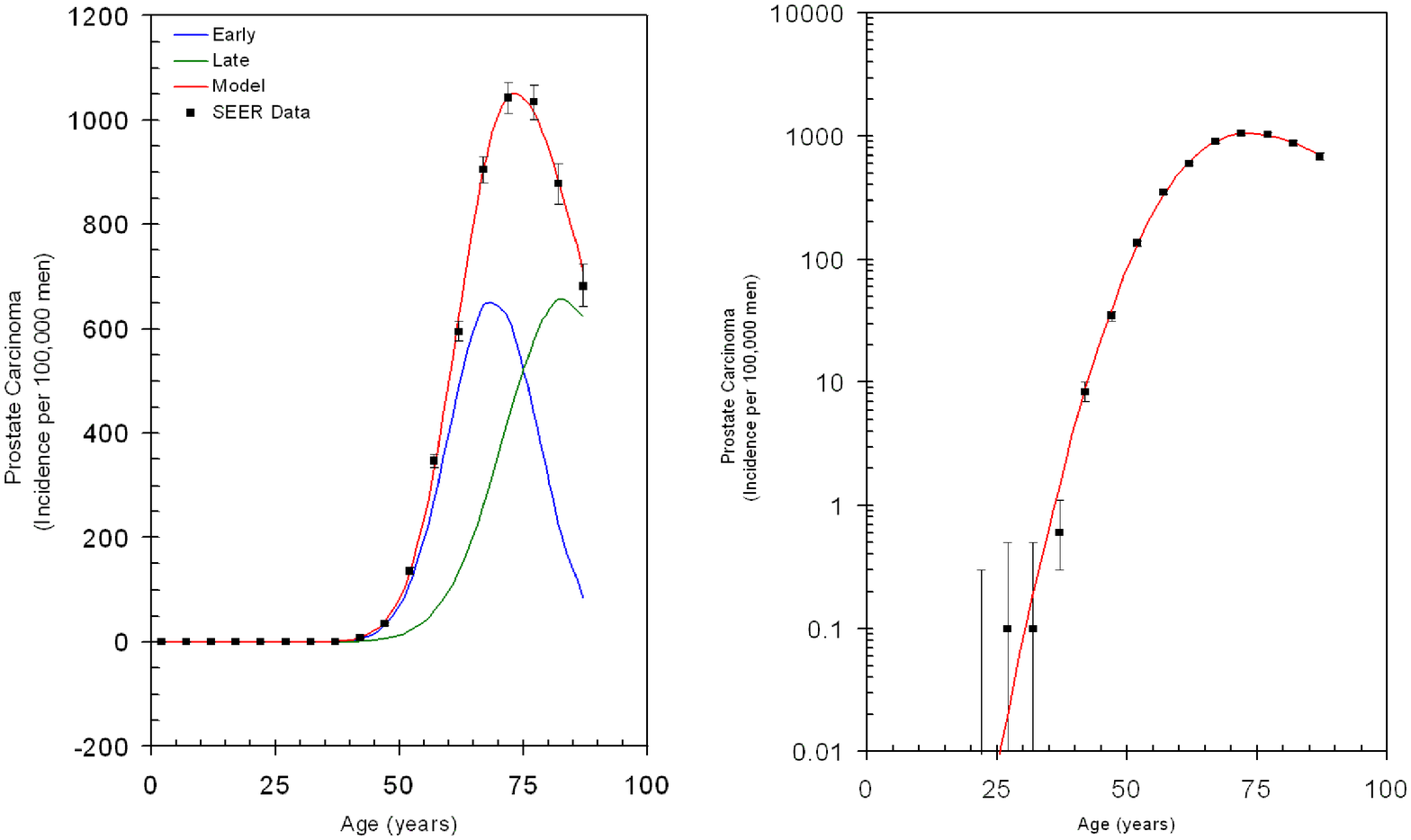}
\end{center}

\caption[]{Supplemental Figure. The age-incidence of prostate
carcinoma. As in Figure 2, the left and right panels show the same
data.  On the left the incidence is plotted on a linear scale, while
on the right it is plotted on a logarithmic scale.  In each case, the
measured incidence is represented by a point and the 95\% confidence
intervals by error bars.  The solid lines represents predicted
incidence levels based upon the model. }

\label{prostate2000}
\end{figure}

\begin{figure}[htb]
\begin{center}
 \includegraphics[width=6in]{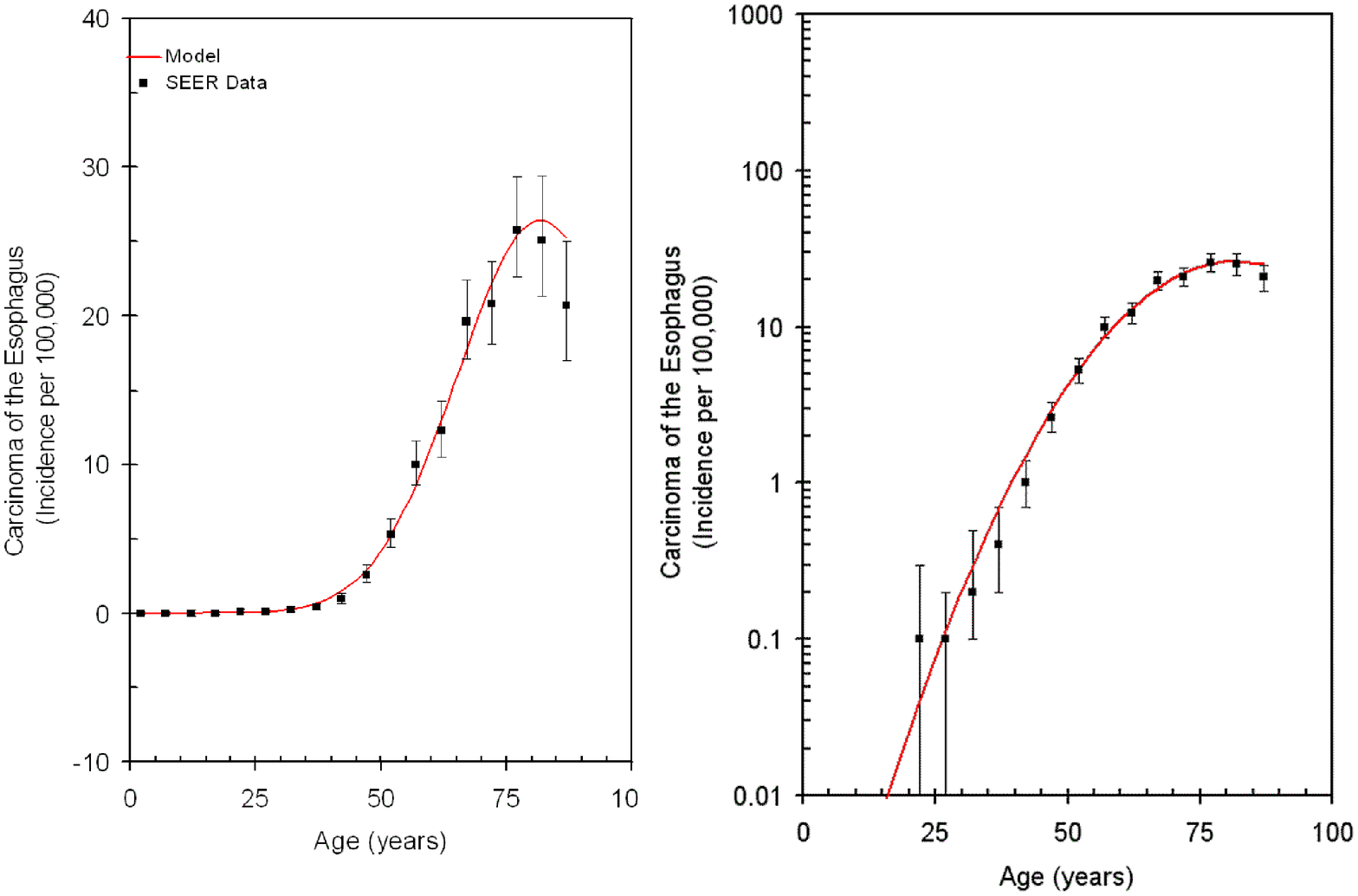}
\end{center}
\caption[]{Supplemental Figure. The age-incidence of carcinoma of the esophagus. As in Figure 2,
the left and right panels show the same data.  On the left the
incidence is plotted on a linear scale, while on the right it is
plotted on a logarithmic scale.    In each case, the measured
incidence is represented by a point and the 95\% confidence intervals
by error bars.  The solid lines represents predicted incidence levels
based upon the  model. }
\label{esophagus2000}
\end{figure}

\begin{figure}[htb]
\begin{center}
 \includegraphics[width=6in]{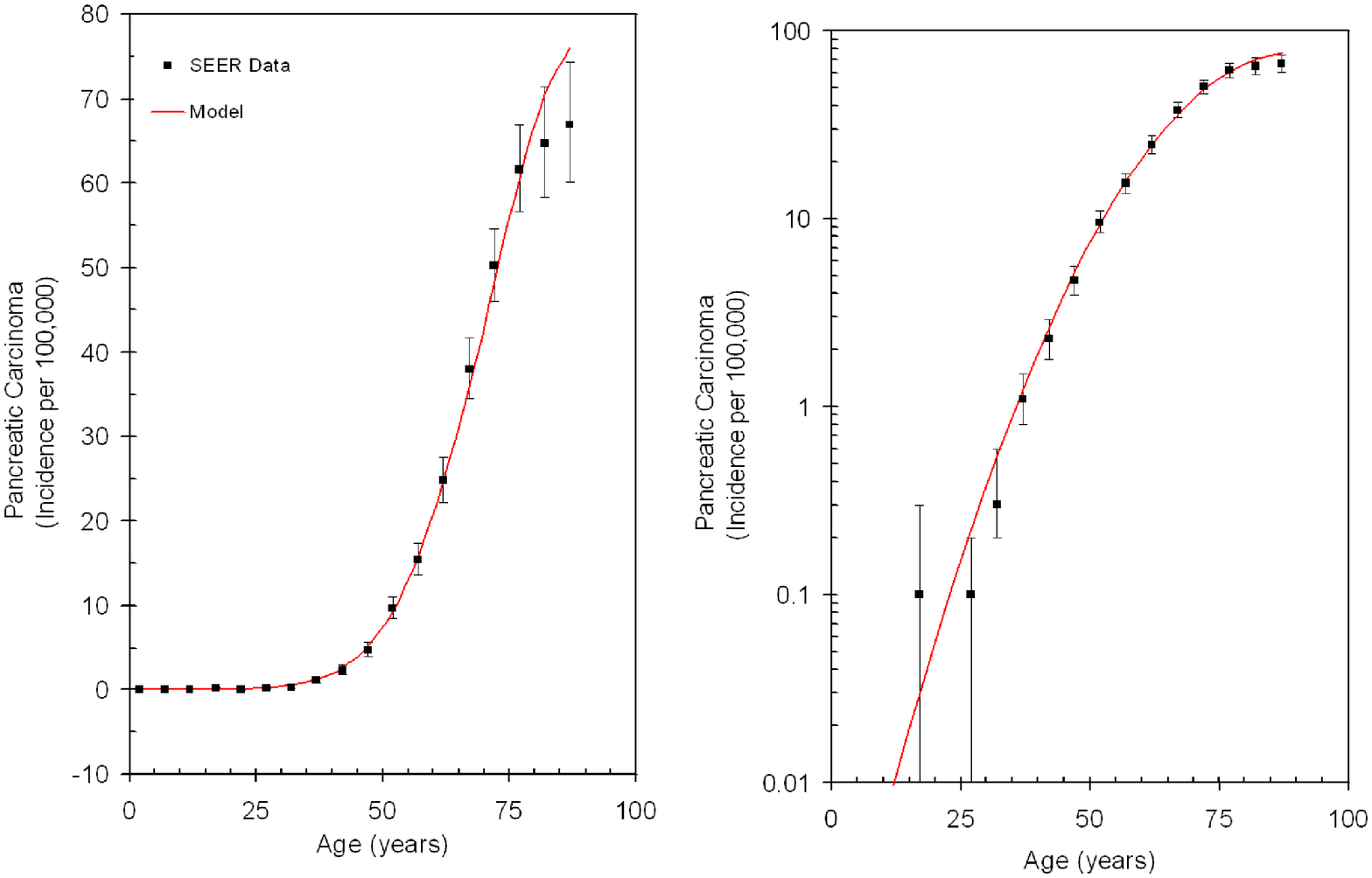}
\end{center}
\caption[]{Supplemental Figure. The age-incidence of carcinoma of the
pancreas. As in Figure 2, the left and right panels show the same
data.  On the left the incidence is plotted on a linear scale, while
on the right it is plotted on a logarithmic scale.  In each case, the
measured incidence is represented by a point and the 95\% confidence
intervals by error bars.  The solid lines represents predicted
incidence levels based upon the model. }
\label{pancreas2000}
\end{figure}

\begin{figure}[htb]
\begin{center}
 \includegraphics[width=6in]{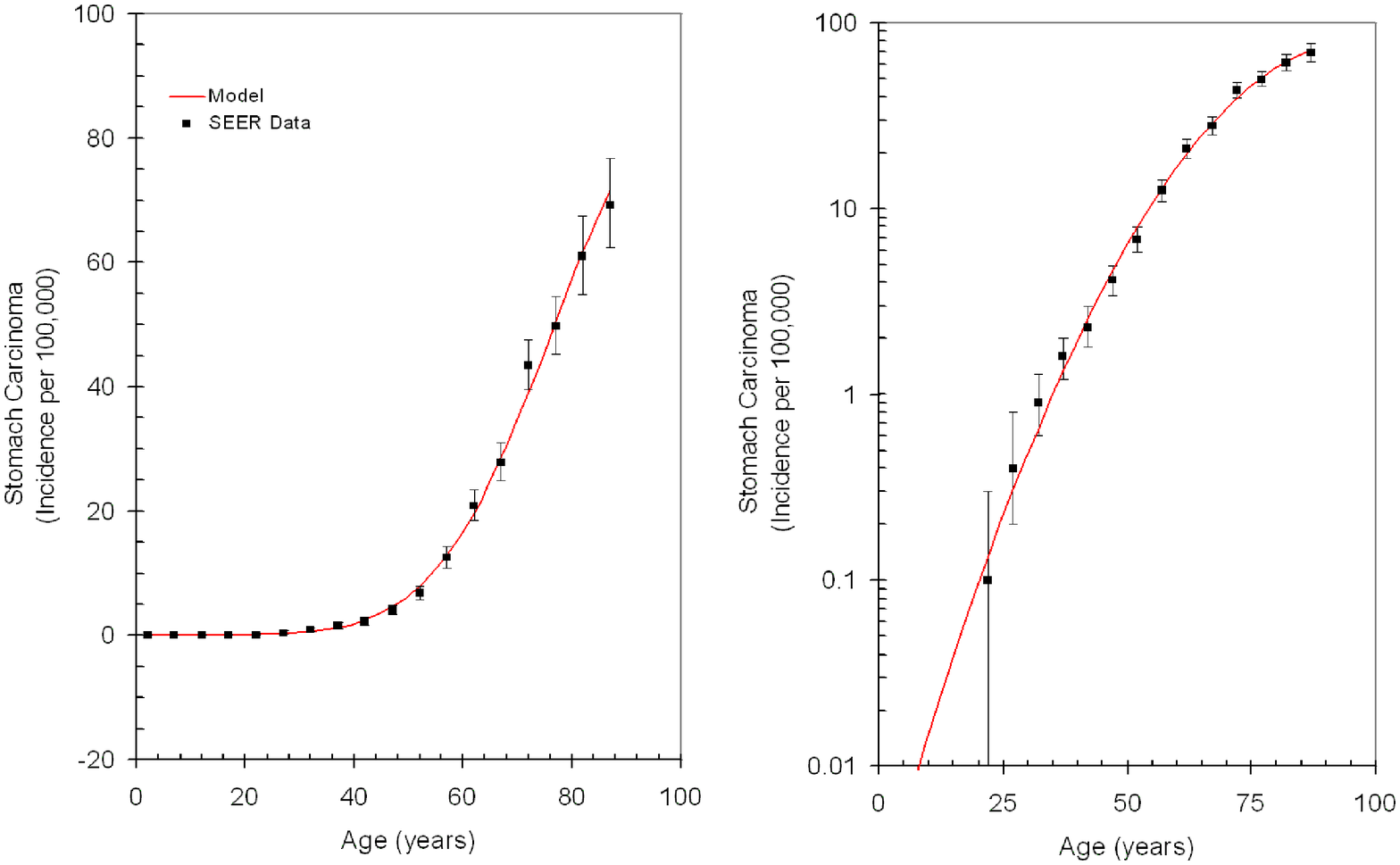}
\end{center}

\caption[]{Supplemental Figure. The age-incidence of carcinoma of the
stomach. As in Figure 2, the left and right panels show the same data.
On the left the incidence is plotted on a linear scale, while on the
right it is plotted on a logarithmic scale.  In each case, the
measured incidence is represented by a point and the 95\% confidence
intervals by error bars.  The solid lines represents predicted
incidence levels based upon the model. }
\label{stomach2000}
\end{figure}

\end{document}